\documentclass[sigplan]{acmart}

\renewcommand\footnotetextcopyrightpermission[1]{} % removes footnote with conference information in first column
\usepackage{booktabs} % For formal tables
\usepackage[utf8]{inputenc} % Encoding
\usepackage{amsfonts}
\usepackage{mathtools}
\usepackage{amsmath}
\usepackage{tabularx}
\usepackage[nomessages]{fp}% http://ctan.org/pkg/fp
\usepackage{geometry}
\usepackage{enumitem}
\usepackage{algpseudocode}
\usepackage{algorithm}
\usepackage{subcaption}
\usepackage{hyperref}
\usepackage{float}
\usepackage{titlesec}
\titleformat{\section}[block]{\sffamily\Large\bfseries\filcenter}{\thesection}{1em}{}

\usepackage{multicol}
\begin{document}
\title{Gelly-Scheduling: Distributed Graph Processing for Service Placement in Community Networks}
\subtitle{\small{This document merely serves the purpose of timely dissemination. Copyrights belong to original holders.}}
%%%%% Helpful pointers to reduce the space taken by acmart author information:
% https://tex.stackexchange.com/questions/349672/how-to-write-names-of-multiple-authors-with-shared-affiliation-in-acm-2017-templ
%%%
\author{Miguel E. Coimbra}
\orcid{0000-0002-7191-5895}
\affiliation{
  \institution{INESC-ID/IST, Universidade de Lisboa}
  \city{Lisbon} 
  \country{Portugal} 
}
\email{miguel.e.coimbra@tecnico.ulisboa.pt}
%%%
\author{Mennan Selimi}
\orcid{0000-0001-9644-5415} 
\affiliation{
  \institution{University of Cambridge}
  \city{Cambridge} 
  \country{UK} 
}
\email{ms2382@cam.ac.uk}
%%%
\author{Alexandre P. Francisco}
\orcid{0000-0003-4852-1641}
\affiliation{
  \institution{INESC-ID/IST, Universidade de Lisboa}
  \city{Lisbon} 
  \country{Portugal} 
}
\email{aplf@inesc-id.pt}
%%%
\author{Felix Freitag}
\orcid{0000-0001-5438-479X}
\affiliation{
   \institution{Universitat Politècnica de Catalunya}
  \city{Barcelona} 
  \country{Spain} 
}
\email{felix@ac.upc.edu}
%%%
\author{Luís Veiga}
\orcid{0000-0002-9285-0736}
\affiliation{
  \institution{INESC-ID/IST, Universidade de Lisboa}
  \city{Lisbon} 
  \country{Portugal} 
}
\email{luis.veiga@inesc-id.pt}
\renewcommand{\shortauthors}{Miguel E. Coimbra et al.}
\renewcommand{\shorttitle}{Gelly-Scheduling}

\makeatletter
\def\@copyrightmode{\relax}
\def\@columnsep{2cm}
\def\@fontsize{15pt}
\def\ACM@fontsize{15pt}
\makeatother

\begin{abstract}
Community networks (CNs) have seen an increase in the last fifteen years.
Their members contact nodes which operate Internet proxies, web servers, user file storage and video streaming services, to name a few.
Detecting communities of nodes with properties (such as co-location) and assessing node eligibility for service placement is thus a key-factor in optimizing the experience of users.
We present a novel solution for the problem of service placement as a two-phase approach, based on: \textit{1)} community finding using a scalable graph label propagation technique and \textit{2)} a decentralized election procedure to address the multi-objective challenge of optimizing service placement in CNs.
Herein we: \textit{i)} highlight the applicability of leader election heuristics which are important for service placement in community networks and scheduler-dependent scenarios; \textit{ii)} present a parallel and distributed solution designed as a scalable alternative for the problem of service placement, which has mostly seen computational approaches based on centralization and sequential execution.
\end{abstract}

\setcopyright{none}

\maketitle

%%%%%%%%%%%%%%%%%%%%%%%%%%%%%%%%%%%%%%%%%%%%%%%%
%%%%%%%%%%%%%%%%%%%%%%%%%%%%%%%%%%%%%%%%%%%%%%%% INTRODUCTION
%%%%%%%%%%%%%%%%%%%%%%%%%%%%%%%%%%%%%%%%%%%%%%%%

\section{Introduction}\label{sec:intro}
Community networks (CNs) are owned and managed by volunteers and offer various services to their members. 
Seamless computing and service sharing in CNs have gained momentum due to the emerging technology of CN micro-clouds.
One such network is guifi.net, located in the Catalonia region of Spain. It is a successful example of this paradigm. Guifi.net is defined as an open, free and neutral CN built by its members pooling resources.
Guifi.net was born in 2004, and until today, has grown into a network of more than 34,000 operational nodes.
Previous work on guifi.net classified services into network-oriented and user-oriented.
For these two types in the Catalonia region, the three most prevalent occurrences were~\cite{selimi2015cloud}: \textit{a) network-oriented services (558 in this region)} -- network graph-servers (39.24\%), DNS servers (35,48\%) and NTP servers (17.20\%); \textit{b) user-oriented services (514 in this region)} -- proxy servers for Internet access (53.50\%), web pages (11.08\%) and communication applications such as VoIP, audio, video and instant messaging (9.33\%).
\begin{figure}
\centering
\includegraphics[width=0.95\linewidth]{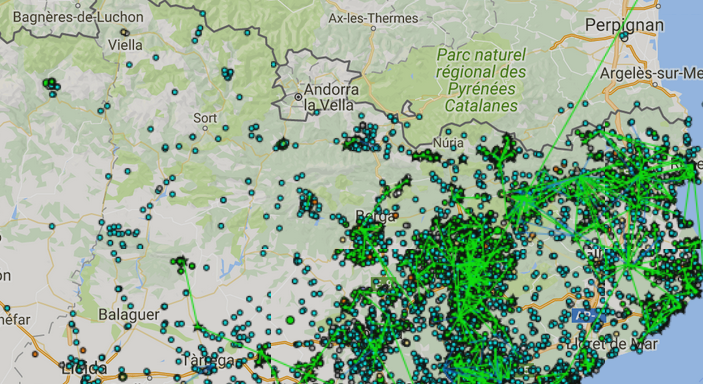}
\caption{Depiction of guifi.net's Osona region.}
\label{fig:guifi-net-map}
\end{figure}
Nodes in guifi.net are exclusive to specific geographical zones (there are no overlays) such as what is depicted in Figure~\ref{fig:guifi-net-map}.
There are special-purpose nodes called graph-servers, which are responsible for performing network measurements between nodes and have an API for querying node states~\cite{MennanSelimi_8_2016}.
These graph-servers comprise a distributed hierarchical monitoring system which records the network's link data traffic properties.
Guifi.net is thus a relevant testbed for developing and validating techniques to enhance service placement and system scheduling by exploring their requirement of leader election.
In turn, these may be extrapolated to more complex scenarios, such as placement in P2P networks (typically irregular), industrial contexts and IoT scenarios.
A simple web proxy would most likely have node latency as its most relevant parameter.
On the other hand, a mission-critical quality-of-service proxy could place the focus on node availability.
Heuristics may encompass network features such as topology, as well as domain attributes (such as availability and quality of specific resources).
While one may intuitively define one heuristic as absolute, this could produce scenarios which are locally optimal but globally undesirable.
What if the node with the highest availability happens to be on the outer rims of the network?
Aspects of network topology are as relevant for system efficiency as the service-level heuristics which traditionally guide leader election for placements.

Our objective is to devise an efficient, scalable solution which is easy to fine-tune regarding domain-specific attributes, and that 
provides seamless scalability for increasing network size and number of services.
For this, we propose a platform that enables incremental processing in a scenario where information continuously arrives: changes in network, node and service quality are continuously monitored.
Our solution is a two-phased approach which optimizes the definition of communities (\textbf{Phase One}) and election of leaders (\textbf{Phase Two}) in a community communications network.
The paper is organized as follows.
Section~\ref{sec:algorithm} explains the two main phases of our algorithm.
Section~\ref{sec:evaluation} details our evaluation methodology and obtained results.
Section~\ref{sec:related} highlights relevant studies on community networks and service placement.
Section~\ref{sec:conclusion} summarizes our contribution's highlights.
%%%%%%%%%%%%%%%%%%%%%%%%%%%%%%%%%%%%%%%%%%%%%%%%
%%%%%%%%%%%%%%%%%%%%%%%%%%%%%%%%%%%%%%%%%%%%%%%% ALGORITHM
%%%%%%%%%%%%%%%%%%%%%%%%%%%%%%%%%%%%%%%%%%%%%%%%
\section{Gelly-Scheduling Service Placement}\label{sec:algorithm}
The challenges inherent to service placement for large scale geo-distributed networks (such as community networks) are usually addressed in the literature with a batch-oriented non-scalable approach. The typical approach consists of performing a search (exhaustive or via heuristics) in a centralized computing unit. All the information about network links and nodes is centrally and sequentially processed, in order to determine the best network configurations as far as service placement is concerned.
While the unit responsible for this search may benefit from hardware improvements, they are merely a form of vertical scaling (which is limited).
This approach does not prioritize reaction to changes in the network and its nodes, in order to make service placement more dynamic in a context of continuous monitoring.
It also doesn't scale in the context of larger networks.
We present a novel method capable of both achieving scale-out processing for optimizing community network topology as well as electing service placement targets within communities in a decentralized approach.
We employ community detection as a parallel technique which enables the partitioning of the problem space to optimize node placement in communities.
This allows for an efficient leader election to execute concurrently (each community being responsible for its leader) and in parallel within each community.
This work aims to improve service placement for networks in a way that users and processing tasks are balanced regarding bandwidth restrictions and data sources.

\begin{figure*}
\centering
\includegraphics[width=0.80\linewidth]{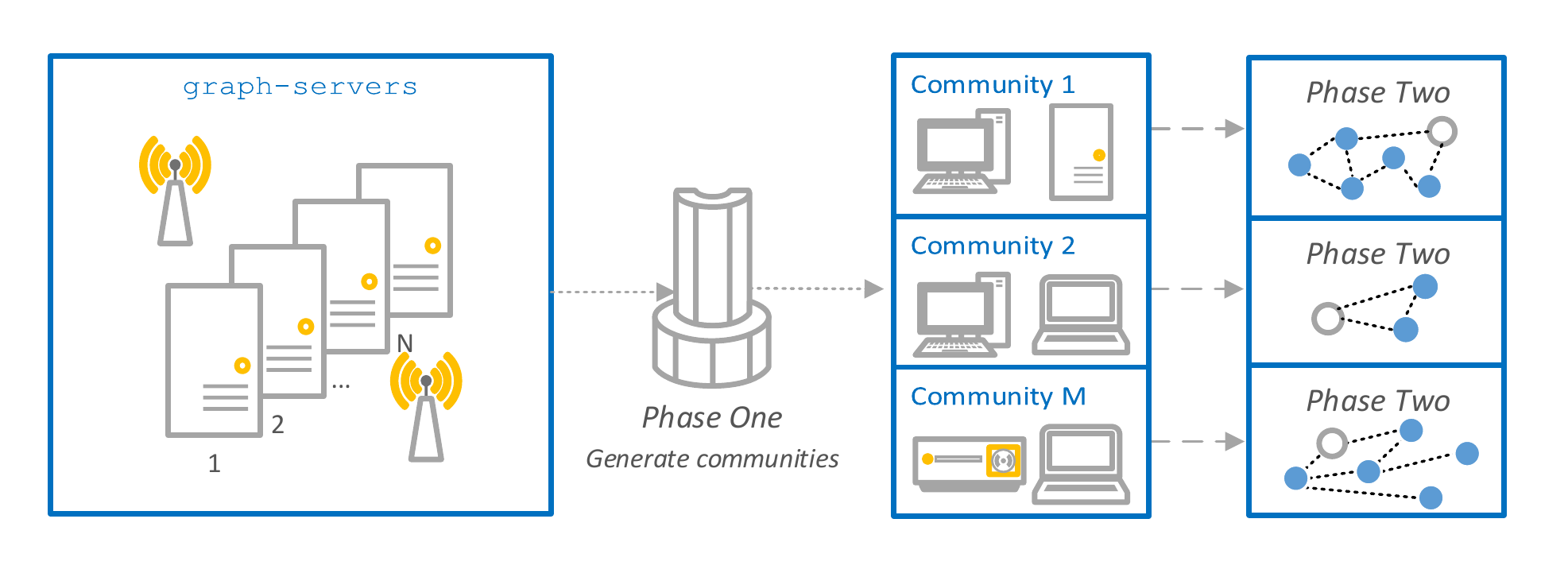}
\caption{The graph-servers on the left send the network heuristics to the master node; the master node in the middle decides on the community configuration through Phase One; communities concurrently elect an internal leader during Phase Two.}
\label{fig:phases}
\end{figure*}

\textbf{Phase One: Community Finding.}
We use two definitions of community: \textit{default --} the zone-based node distributions, provided in the dataset as-is (insignificant preprocessing is performed in this case); \textit{custom --} a state-of-the-art label propagation technique~\cite{PhysRevE.76.036106} applied for detecting communities.
We build an undirected graph $G=(V,E)$ by defining a set of $n$ nodes $V$ and a set of $m$  edges $E$ such that an edge $e \in E$ will be created if and only if there is a corresponding link element between two working devices (each belonging to a working node) in the dataset.
Single-leaf nodes were discarded as part of preprocessing.
The goal of this phase is to rapidly partition the problem space into a configuration that promotes scalability of computation and efficient resource usage.
We provide the pseudo-code for the most relevant actions of Phase One in Algorithm~\ref{alg:one}, where $C$ is an upper bound on the number of iterations to execute (a default limit of $C = 10$ iterations is common in the literature for convergence~\cite{Boldi:2011:LLP:1963405.1963488}).
\begin{algorithm}[b]
\floatname{algorithm}{Algorithm}
\caption{Phase One: Community Finding}
\label{alg:one}
\begin{algorithmic}[1]
\State \textbf{INPUT:} $G = $($V, E$)$, C = 10$
\State \textbf{OUTPUT:} $Z$ \Comment{Set of graphs representing communities}
\ForAll{$v \in V$}
\State $v$.generateUniqueLabel()
\EndFor
\State $G' \longleftarrow G$.setUndirectedEdges()
\State $i \longleftarrow 1$
\For{$i < C$}
\ForAll{$v \in V$}
\State $M \longleftarrow v$.getInboundMessages()
\State $L \longleftarrow M$.getMostFrequentLabels()
\State $v$.updateLabel($L$.filterHighestLabel())
\EndFor
\State $i \longleftarrow i + 1$
\If{\textbf{not} $G'$.labelsChanged()}
\State break
\EndIf
\EndFor\\
\Return $Z
\longleftarrow G'$.groupByLabels()
\end{algorithmic}
\end{algorithm}
Phase One thus becomes an important instrument in efficiently defining groups of network nodes by employing a state-of-the-art technique in community detection.
These groups aid the optimization process of service placement, effectively serving as a useful blueprint for Phase Two of our algorithm.
The two phases form a technique to harness current platforms and infrastructures to tackle service placement.
Conceptually, there is a top-tier master node which is responsible for: \textit{1)} querying the graph-servers for all of the network's node information; \textit{2)} executing Phase One of our algorithm to obtain a definition of communities; \textit{3)} informing each node of its community's composition.
This is depicted in Figure~\ref{fig:phases}, where in the middle there is a centralized entity consisting of one or (potentially many) more computational workers.
It initially queries graph-servers (or whatever network visibility mechanisms are in place) to obtain a snapshot of the network's nodes.
Then it executes Phase One of our algorithm, decomposing the network into communities.
A major computational advantage of Phase One is that this master can be a single machine or a set of workers in a cluster, effectively scaling with the computational capability available to this top-tier master.
Each community member is then informed of the elements of its own community:  required to proceed to Phase Two.

\textbf{Phase Two: Leader Election.}
Phase Two receives a set of communities and elects a leader for each one.
This election phase is self-contained for each community, in the sense that a distributed implementation of this phase can be carried out concurrently with respect to communities and in parallel within each community with our graph-based approach.
The right-side of Figure~\ref{fig:phases} illustrates this.
There may be more than one connected component in geographical zones of guifi.net.
Due to this, for every community network $G$, only the nodes belonging to the largest connected component of $G$ are used to choose a leader for service placement.
This election consists of Phase Two of our algorithm and is detailed in Algorithm~\ref{alg:two}.
This phase serves the purpose of identifying the best node for service placement.
Leadership is attributed through a scoring, where the score of each node $i$ lies in defining a linear combination of two sets of heuristics.
One set is based on system-centric values: availability $\beta_{1}$ and latency $\beta_{2}$ as defined by graph-servers~\cite{6379103}, as well as computational class $\beta_{3}$ as per Table~\ref{table:node-categories} and defined as part of this work; the other is calculated as part of this algorithm and consists of betweenness $\alpha_{1}$ and closeness $\alpha_{2}$ centralities.
We defined heuristic $\beta_{3}$ as a score in three computational categories for nodes: \textit{i)} server-type nodes which typically have stronger computational power to support more demanding services;
\textit{ii)} non-server nodes with more than one device; \textit{iii)} non-server nodes with a single device.
Table~\ref{table:node-categories} shows the representation of $\beta_{3}$ for each category for the data we analyzed.
The values we attribute to $\beta_{3}$ were selected arbitrarily to represent computational power of a given node.
\begin{table}[t]
\centering
%http://tex.stackexchange.com/questions/7208/how-to-vertically-center-the-text-of-the-cells#7318
  \caption{Frequency of per-node device count categories. The most frequent services are Internet proxies, a consequence of guifi.net existing as an alternative to the standard ISP model.}
    \label{table:node-categories}
    \begin{tabular}{|>{\centering\arraybackslash} m{1.70cm}|>{\centering\arraybackslash} m{2.45cm}|>{\centering\arraybackslash} m{0.60cm}|}
    \hline
    Nodes &  23,468 (100\%) &  $\beta_{3}$ \\ \hline
    \textit{i)} Strong &  337 (1.436\%) & 1 \\ \hline
    \textit{ii)} Medium & 1,666 (7.099\%)  & 0.5 \\ \hline
    \textit{iii)} Weak &  21,465 (91.465\%) & 0.1 \\ \hline
    \end{tabular}
\end{table}
This categorization serves the purpose of approximating realistic tiers of computational capabilities for nodes in the network -- information which, as far as the authors know, is not readily-available in the guifi.net CNML dataset.
Thus, as an example, the initial score of a node $i$ will be defined as:
\begin{equation}
s_{i} = w_{1}\ \alpha_{1} + w_{2}\ \alpha_{2} + w_{3}\ \beta_{1} + w_{4}\ \beta_{2} + w_{5}\ \beta_{3}
\end{equation}
Table~\ref{table:heuristics} details the specifics of each heuristic, namely their meaning and how they are obtained.
Notation-wise, $i$ is the node to be scored while $u$ and $v$ represent arbitrary nodes in the community graph $G$ with $n$ nodes, $\sigma_{u, v}$ is the number shortest paths from $u$ to $v$, $\sigma_{u, v}(i)$ is the number of those that pass through $i$, and $d(i, v)$ is the geodesic distance between $i$ and $v$.
Phase Two was designed under two types of evaluation based on configuration of heuristics:
\textbf{Absolute heuristics} - in this case, leader selection is guided exclusively by exactly one of the heuristics.
We analyze the impact of each individual heuristic, setting the weights of others to zero.
\textbf{Combined heuristics} - we consider a linear combination of two heuristics.
We set unbalanced weights in order to better determine the more significant contributions, in the sense that for two heuristics $m_{1}$ and $m_{2}$, we may define the node score to be $v_{s} = (1 - f)m_{1} + fm_{2}$, or the reverse.
If one heuristic weights in for 60\% of the score, the other will account for the remaining 40\%.
\begin{algorithm}[t]
\floatname{algorithm}{Algorithm}
\caption{Phase Two: Leader Election}
\label{alg:two}
\begin{algorithmic}[1]
\State \textbf{INPUT:} $G = $($V, E$)$, W$ \Comment Heuristic weight array
\State \textbf{OUTPUT:} $R$ \Comment{Decreasing-order ranks}
\State $\alpha \longleftarrow $[ ], $\beta \longleftarrow$[ ]
\ForAll{$v \in V$}
\State $\alpha$[$v$]$ \longleftarrow v$.calculateAlphas()
\State $\beta$[$v$]$ \longleftarrow v$.getBetas()
\State $v$.setScore($W \ * \ $[\ $\alpha$[$v$]$\quad \beta$[$v$]\ ])
\EndFor
\State $R \longleftarrow G$.getVertices().orderByScore()\\
\Return $R$
\end{algorithmic}
\end{algorithm}

\newcommand{\specialcell}[2][c]{%
    \begin{tabular}[#1]{@{}l@{}}#2\end{tabular}
}
\begin{table}[htbp]
\centering
\caption{Algorithm's heuristic symbols and meanings.}
\label{table:heuristics}
\begin{tabularx}{\textwidth}{p{0.05\columnwidth} | p{0.85\columnwidth}}
$\alpha_{1}$& \specialcell{\textbf{Betweenness Centrality}\\ \( \sum_{u\neq i \neq v}\frac{\sigma_{u v}(i)}{\sigma_{u v}} \), fraction of shortest paths from $u$ to\\
$v$, for all nodes $u$ and $v$, passing through node $i$.} \\
$\alpha_{2}$& \specialcell{\textbf{Closeness Centrality}~\cite{Newman:2010:NI:1809753}\\  $(n - 1)/\sum_{v}{d(i, v)}$, where $d(v,i)$ is the geodesic\\ distance from node $i$ to  node $v$.} \\
$\beta_{1}$& \specialcell{\textbf{Availability}\\Percentage of ping responses received by a graph-\\-server (\%) over a specific time period.} \\
$\beta_{2}$& \specialcell{\textbf{Latency}\\Ping response timing, measured by a graph-server\\ (ms) over a specific time period.} \\
$\beta_{3}$& \specialcell{\textbf{Computational Class}\\Defined by the number of devices handled by the \\node, as well as its role.} 
\end{tabularx}
\end{table}

%%%%%%%%%%%%%%%%%%%%%%%%%%%%%%%%%%%%%%%%%%%%%%%%
%%%%%%%%%%%%%%%%%%%%%%%%%%%%%%%%%%%%%%%%%%%%%%%% IMPLEMENTATION DETAILS
%%%%%%%%%%%%%%%%%%%%%%%%%%%%%%%%%%%%%%%%%%%%%%%%
\subsection{Implementation}\label{sec:sec:implementation}

Scores of heuristics $\alpha_{1}$ and $\alpha_{2}$ were obtained for each community $G$ using the \texttt{Python NetworkX} library, for use in Phase Two.
Overall, the time to calculate them is negligible when compared to the total amount of time required to compute Phase One plus Phase Two.
There are common aspects to generating samples for bandwidth and round-trip time, but each was based on different statistical artifices.

\textbf{Bandwidth.}
Let $BW$\ \textasciitilde\ $K$($k, h, \xi, \alpha$) represent the empirical bandwidth distribution.
$K$ stands for the four-parameter Kappa distribution~\cite{hosking1994four}, where $k$ and $h$ denote the shape of the distribution, $\xi$ denotes its location and $\alpha$ is a scaling factor.
These four parameters were estimated using L-moment statistics, namely through the \texttt{lmoms} function which computes the sample L-moments and the \texttt{parkap} function which estimates the four parameters of $K$ based on the sample L-moments.
Both functions are part of the \texttt{R lmomco} library.
The four-parameter Kappa distribution is used for simulating additional samples based on the empirical distribution made by using the \texttt{rkappa4} function of the \texttt{R FAdist} library for random generation purposes.

\textbf{Round-trip time.}
Let $RTT$\ \textasciitilde\ $GEV$($\mu, \sigma, \xi$) represent the empirical round-trip time distribution.
$GEV$ is the generalized extreme value distribution.
It has four parameters: $\mu$, which is the location of the distribution, $\sigma$ which represents the scale and $\xi$ which represents the shape of $GEV$ (influencing the behavior of the distribution tail).
The previously-referenced \texttt{lmoms} function was used as well, with \texttt{pargev} now being the function (also present in library \texttt{lmomco}) responsible for estimating the $GEV$ parameters based on the sample L-moments.
Additional round-trip time samples were simulated using the \texttt{rgev} function.
$GEV$ exists as a family of continuous probability distributions, stemming from extreme value theory~\cite{coles2001introduction}.

\begin{figure}
\centering
\includegraphics[width=0.950\linewidth]{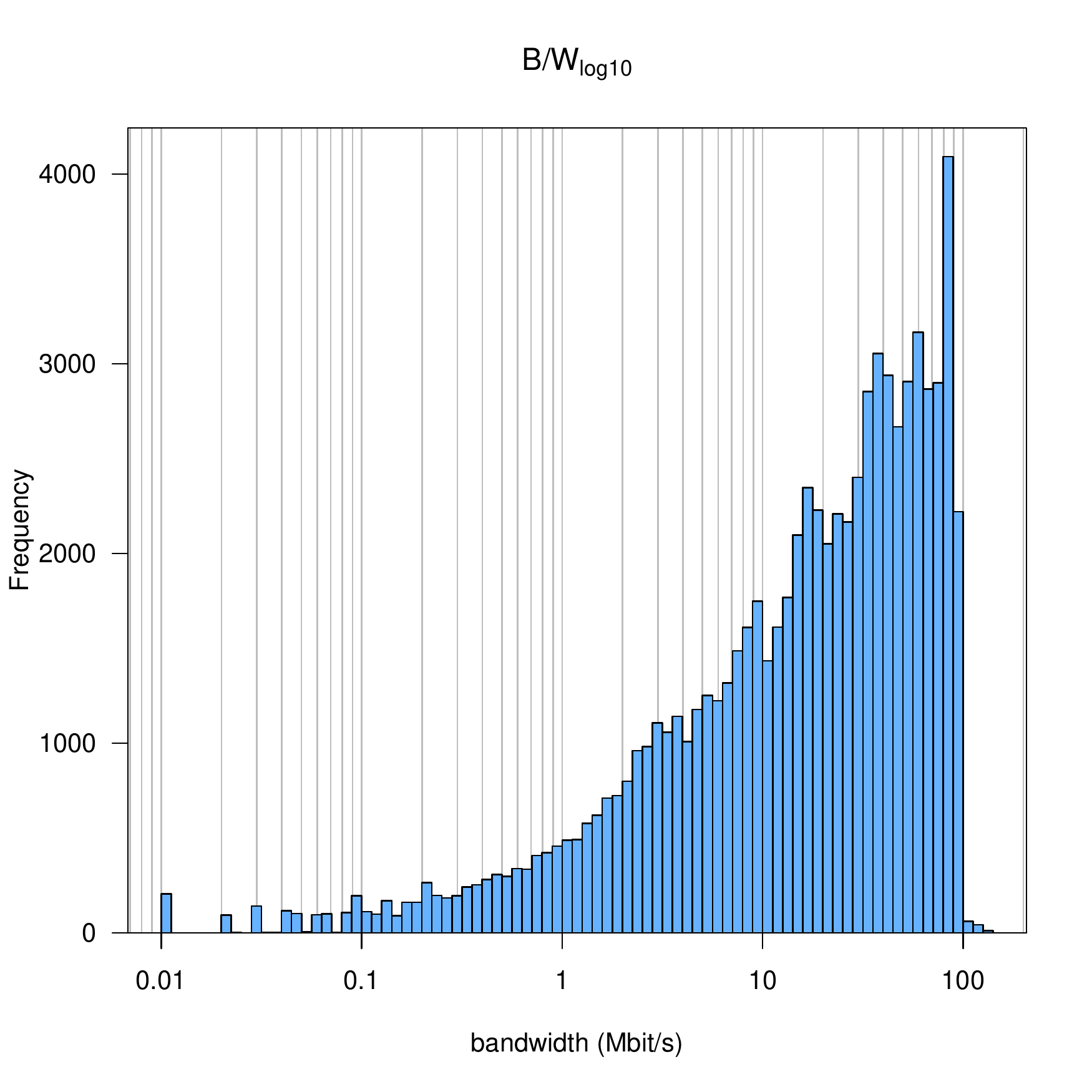}
\caption{Bandwidth (B/W) in logarithmic scale.}
\label{fig:bw-log}
\end{figure}

\begin{figure}
\centering
\includegraphics[width=0.950\linewidth]{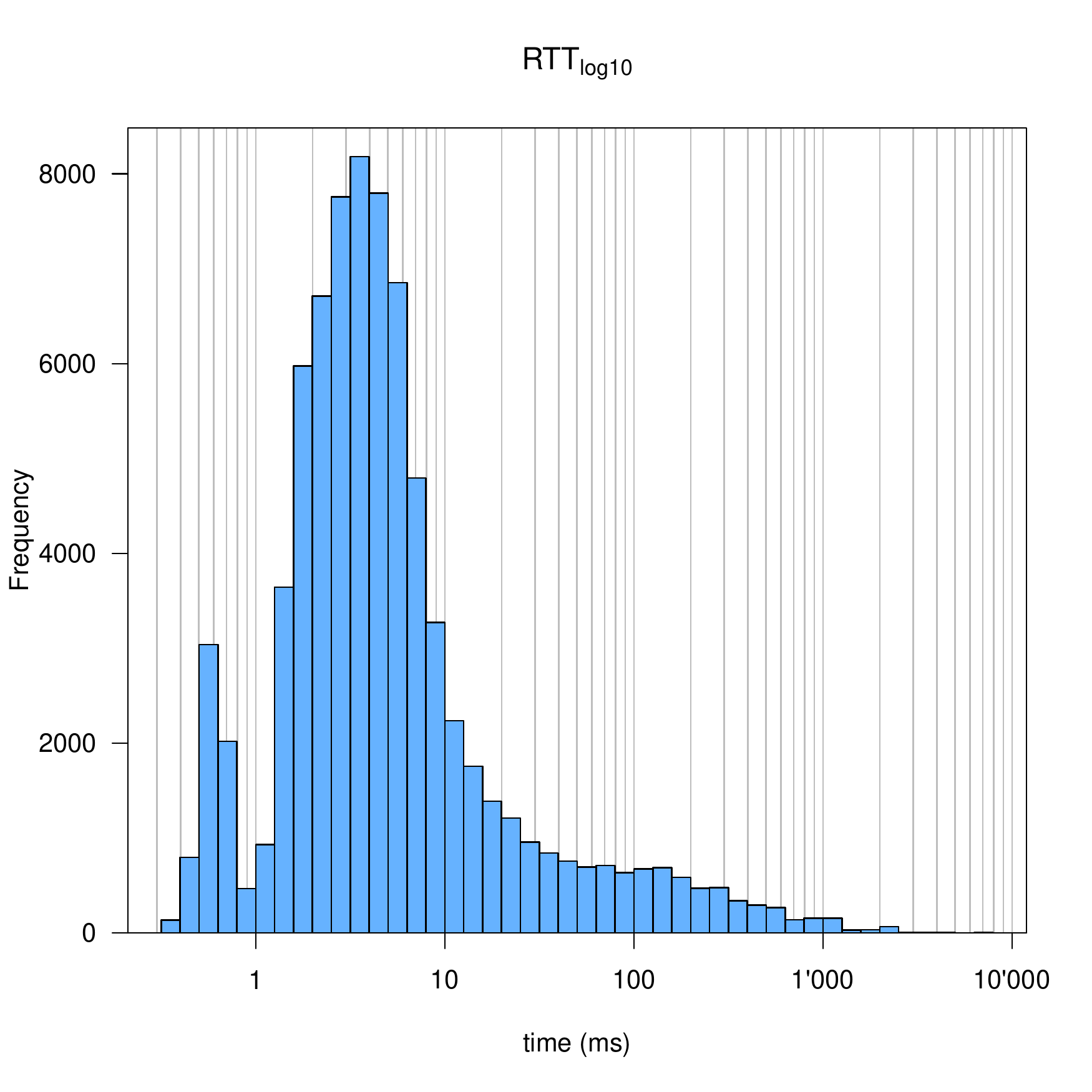}
\caption{Round-trip time (RTT) in logarithmic scale.}
\label{fig:rtt-log}
\end{figure}

The method of L-moments is used to understand insights of analyzed data and to estimate distributions~\cite{10.2307/2345653,hosking2005regional} using efficient techniques~\cite{hosking2000fortran}.
Figure~\ref{fig:bw-log} shows the $\log_{10}$ plot of the bandwidth, while Figure~\ref{fig:rtt-log} shows the same for round-trip time.


%%%%%%%%%%%%%%%%%%%%%%%%%%%%%%%%%%%%%%%%%%%%%%%%
%%%%%%%%%%%%%%%%%%%%%%%%%%%%%%%%%%%%%%%%%%%%%%%% EVALUATION
%%%%%%%%%%%%%%%%%%%%%%%%%%%%%%%%%%%%%%%%%%%%%%%%

\section{Experimental Evaluation}\label{sec:evaluation}

\begin{figure}
\centering
\includegraphics[width=0.95\linewidth]{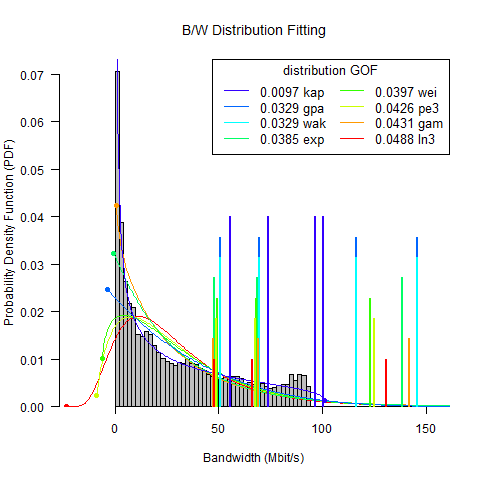}
\caption{Bandwidth (B/W) observation comparison and goodness-of-fit of different candidate distributions.
Lower goodness-of-fit is better.}
\label{fig:bw-fits}
\end{figure}

\begin{figure}
\centering
\includegraphics[width=0.95\linewidth]{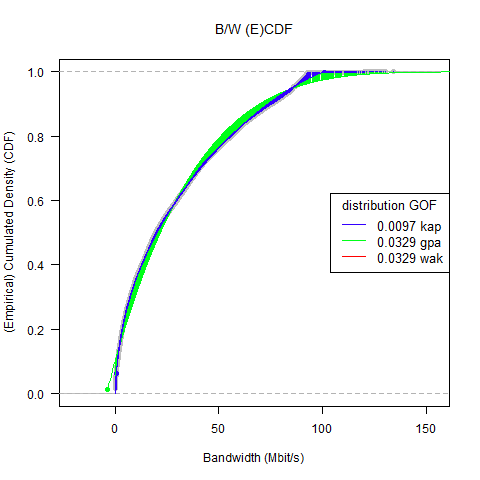}
\caption{Bandwidth (B/W) Top 3 Empirical Cumulative Distribution Function (ECDF).}
\label{fig:bw-ecdf}
\end{figure}

\begin{figure}
\centering
\includegraphics[width=0.95\linewidth]{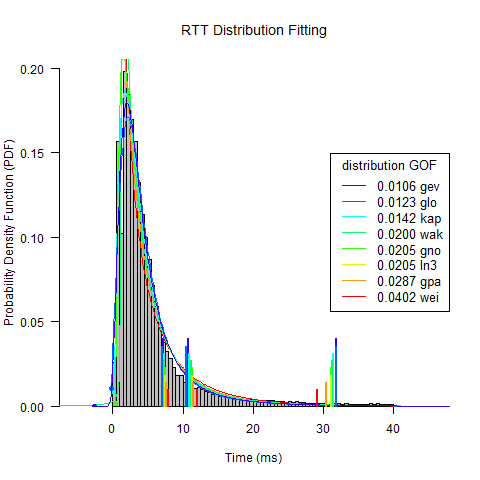}
\caption{Round-trip time (RTT) observation comparison and goodness-of-fit of different candidate distributions.
Lower goodness-of-fit is better.}
\label{fig:rtt-fits}
\end{figure}

\begin{figure}
\centering
\includegraphics[width=0.95\linewidth]{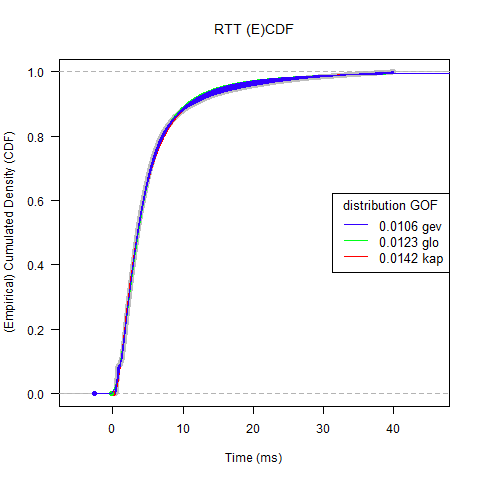}
\caption{Round-trip time (RTT) Top 3 Empirical Cumulative Distribution Function (ECDF).}
\label{fig:rtt-ecdf}
\end{figure}

There are 23,391 nodes identified as working and, for the whole guifi.net, there are 878 nodes defined as servers.
This implies that, at most, 3.75\% of the working nodes could actually be sustaining full fledged services.
We believe guifi.net, while it is in fact an open community network, has a type of topology which allows for extrapolating results into other sorts of networks.
This claim is made based on previous research work in the literature~\cite{MennanSelimi_8_2016}, which both analyzed the impact of prioritizing different heuristics on the computational and network resources available~\cite{MennanSelimi_8_2016_2} and studied practical issues with micro-service architectures~\cite{MennanSelimi_4_2017}.
We used available statistical processing tools to attempt to fit several distributions and compare them.
For the bandwidth, we present plots of the distribution fitting and Empirical Cumulative Distribution Function in Figures~\ref{fig:bw-fits} and~\ref{fig:bw-ecdf}.
In the same order, we also present the aforementioned plots for round-trip time in Figures~\ref{fig:rtt-fits} and~\ref{fig:rtt-ecdf}.

We then modeled the ECDF of both network properties with the use of the \texttt{lmomco} and \texttt{FAdist} libraries in \texttt{R}.

\textbf{Network Characteristics.}
Part of guifi.net exists as an instance of the \textit{Quick Mesh Project} (QMP\footnote{http://qmp.cat/}), a system for easily deploying MESH/MANET networks using Wi-Fi technology. QMP is an urban mesh network in Barcelona and it is a subset of the guifi.net community network sometimes called Sants-UPC network. It was designed for use in scenarios such as free community networks, of which guifi.net is a rich example~\cite{MennanSelimi_4_2017}.
We use measurements of round-trip time (RTT) and bandwidth (B/W) from the Sants-UPC wireless mesh QMP instance to establish a model of these telecommunication heuristics for the remainder of the network.
It would be through a hierarchy of graph-server nodes that one would acquire a view of all the nodes in the network.
However, due to privacy and maintenance issues, many of these graph-server types fail to provide any type of information about queried nodes.
Due to this, we employed a one-week snapshot of this seventy-node QMP instance to establish ground-truth relevance for our work.
%The measurements were taken for seven days from the 1st to the 10th of March 2017, with a snapshot taken every hour~\cite{6379103}.
The measurements were taken for seven days from the 1st to the 8th of March 2017, with a snapshot taken every hour~\cite{6379103}.
The measurement period and frequency produced enough samples for evaluating guifi.net in light of the results of our method.
We remark that node links in QMP and guifi.net, in general, are not symmetrical: the bandwidth and round-trip time from node $u$ to node $v$ isn't necessarily the same from $v$ to $u$.

\begin{figure*}
\centering
\includegraphics[width=0.95\linewidth]{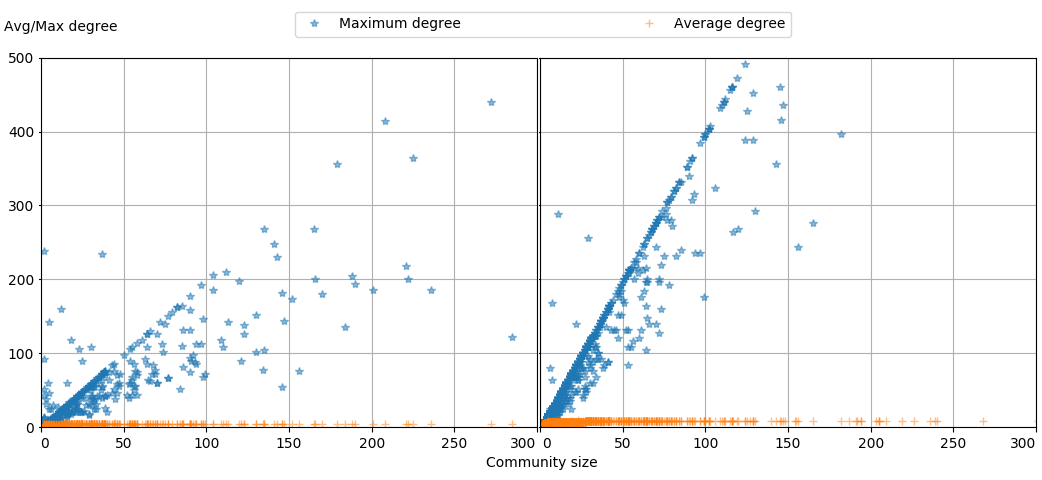}
\caption{Plot of maximum and average degree distributions for each community. The left image is the geographical configuration of node sets, while the right side is based on Phase One of our algorithm.}
\label{fig:avg-degrees}
\end{figure*}

\begin{figure*}
\centering
\includegraphics[width=0.95\linewidth]{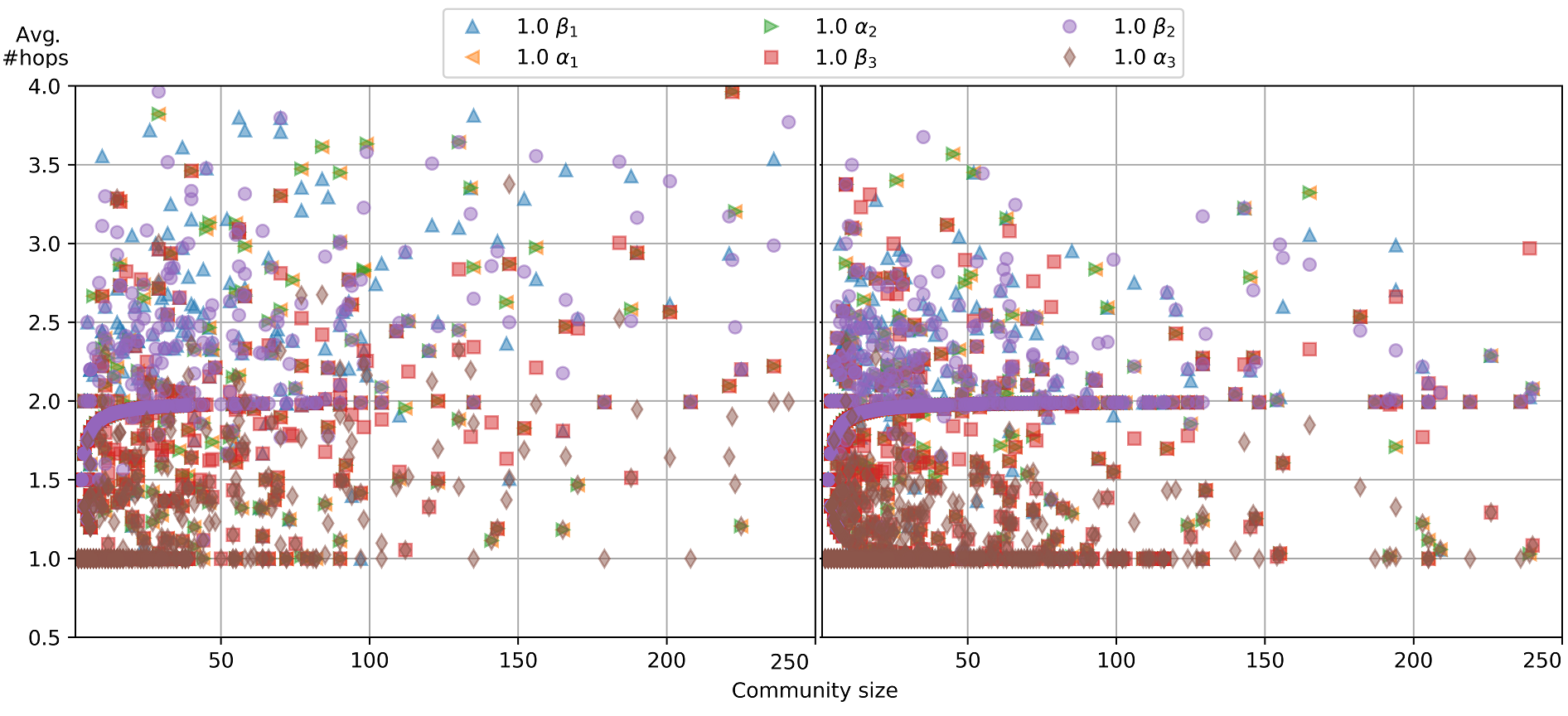}
\caption{Average number of hops-to-leader plotted against each community's size. The left image is the geographical configuration of node sets, while the right side is based on Phase One of our algorithm.}
\label{fig:avg-hops}
\end{figure*}

Firstly, although QMP has a more uniform set of nodes compared to guifi.net, it is also subject to the same behavioral user factors which influence the whole network~\cite{6379139}.
This means that it may be considered as a representative sampling of guifi.net.
Secondly, the obtained number of samples is high enough to enable us to apply statistical techniques to define empirical models of bandwidth and round-trip time.
This allows us to fit different distributions to the measurements and evaluate the resulting goodness-of-fit (GOF) values.
Selecting the most fitting distributions, we then synthesize their parameters in order to generate functions to produce artificial observations.
Qualitatively, these simulated values are representative of the behavior of the QMP network (and thus of guifi.net) and were used to populate the bulk of our dataset (guifi.net snapshot of January, 2017) nodes, which were missing data.
%Based on results in~\cite{2015:TOG:2852375.2852741} (\textit{Empirical Cumulative Distribution Function} of node availability), experiments were performed with availability generated under distribution $\beta_{1} \sim \mathcal{N}(0.8, 0.2)$ and latency under (heterogeneity in guifi.net is such that urban areas have typically lower latencies and rural ones have increased latencies) distribution $\beta_{2} \sim \mathcal{N}(200, 200)$ (ms).

%%%%%%%%%%%%%%%%%%%%%%%%%%%%%%%%%%%%%%%%%%%%%%%%
%%%%%%%%%%%%%%%%%%%%%%%%%%%%%%%%%%%%%%%%%%%%%%%% EVALUATION - PHASE ONE
%%%%%%%%%%%%%%%%%%%%%%%%%%%%%%%%%%%%%%%%%%%%%%%%
\textbf{Phase One: Network Impact.}
We present in Figure~\ref{fig:avg-degrees} the distribution of maximum and average degree versus the size of the communities.
The left side pertains guifi.net zone-based communities (from the dataset as-is), while the right side is related to the configuration of network node groups obtained with Phase One of our algorithm.
We derive from this that our algorithm produces groupings with a tendency for greater node inter-connectivity.

\begin{figure*}
\centering
\includegraphics[width=0.95\linewidth]{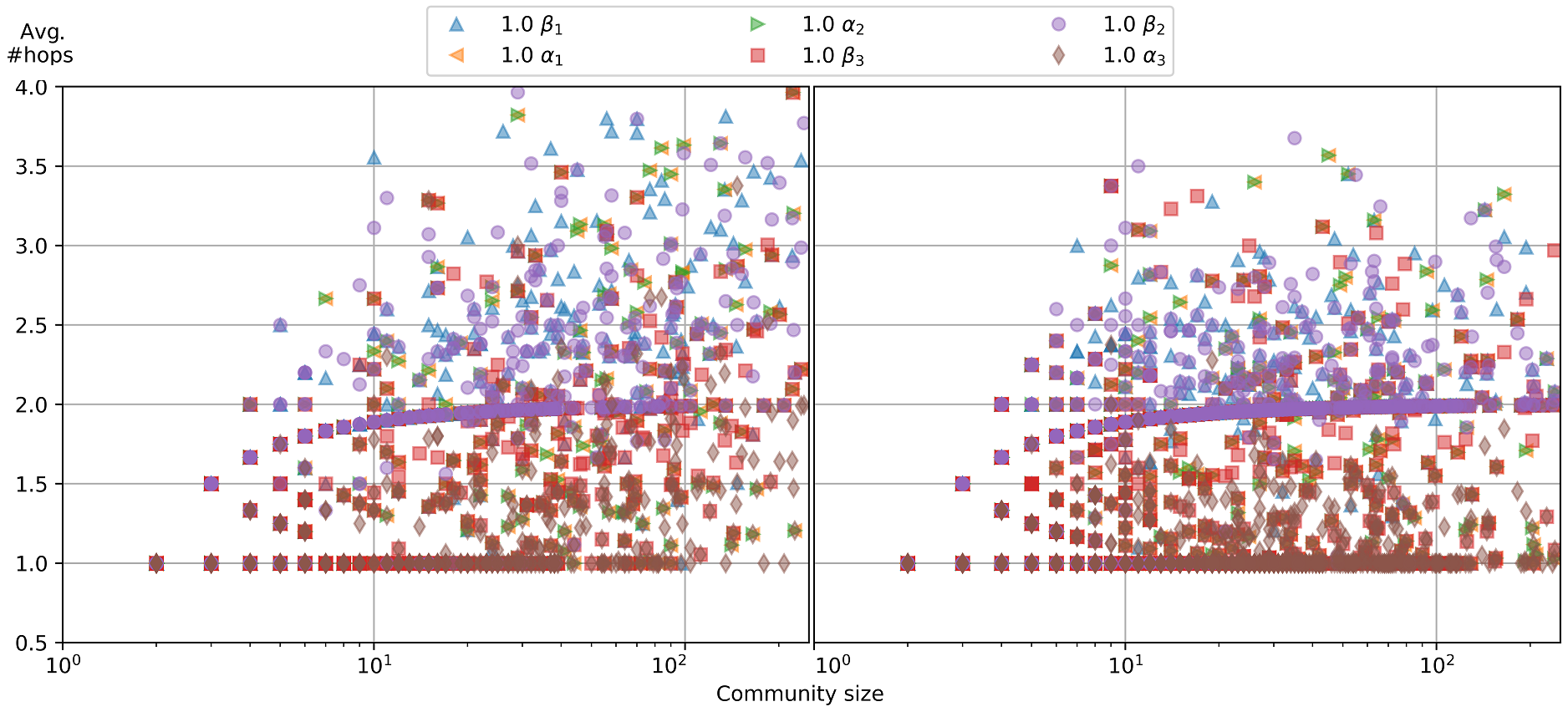}
\caption{Average number of hops-to-leader plotted against a logarithmic scale of each community's size. The left image is the geographical configuration of node sets, while the right side is based on Phase One of our algorithm.}
\label{fig:avg-hops-log}
\end{figure*}

\begin{figure}
\centering
\includegraphics[width=0.95\linewidth]{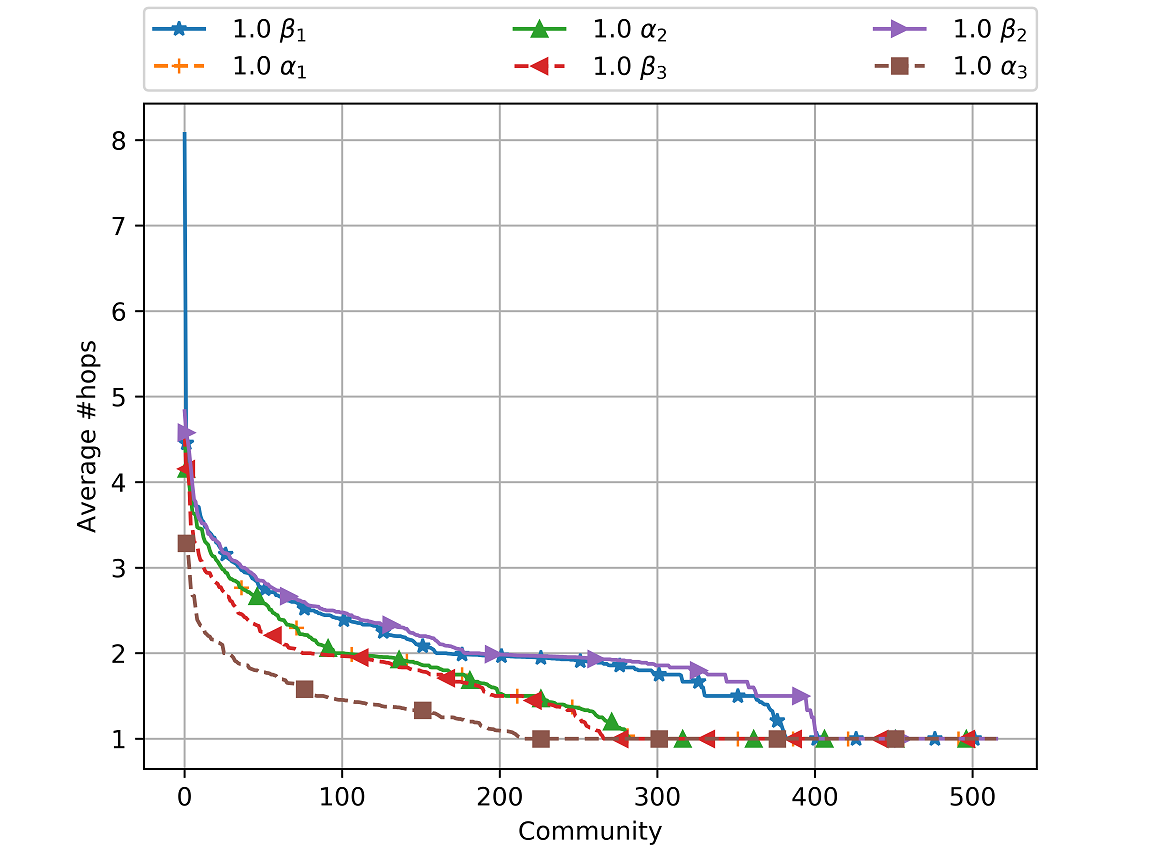}
\caption{Average number of hops-to-leader against community in decreasing order for the original geographical configuration.}
\label{fig:dual-hops-geo}
\end{figure}

\begin{figure}
\centering
\includegraphics[width=0.95\linewidth]{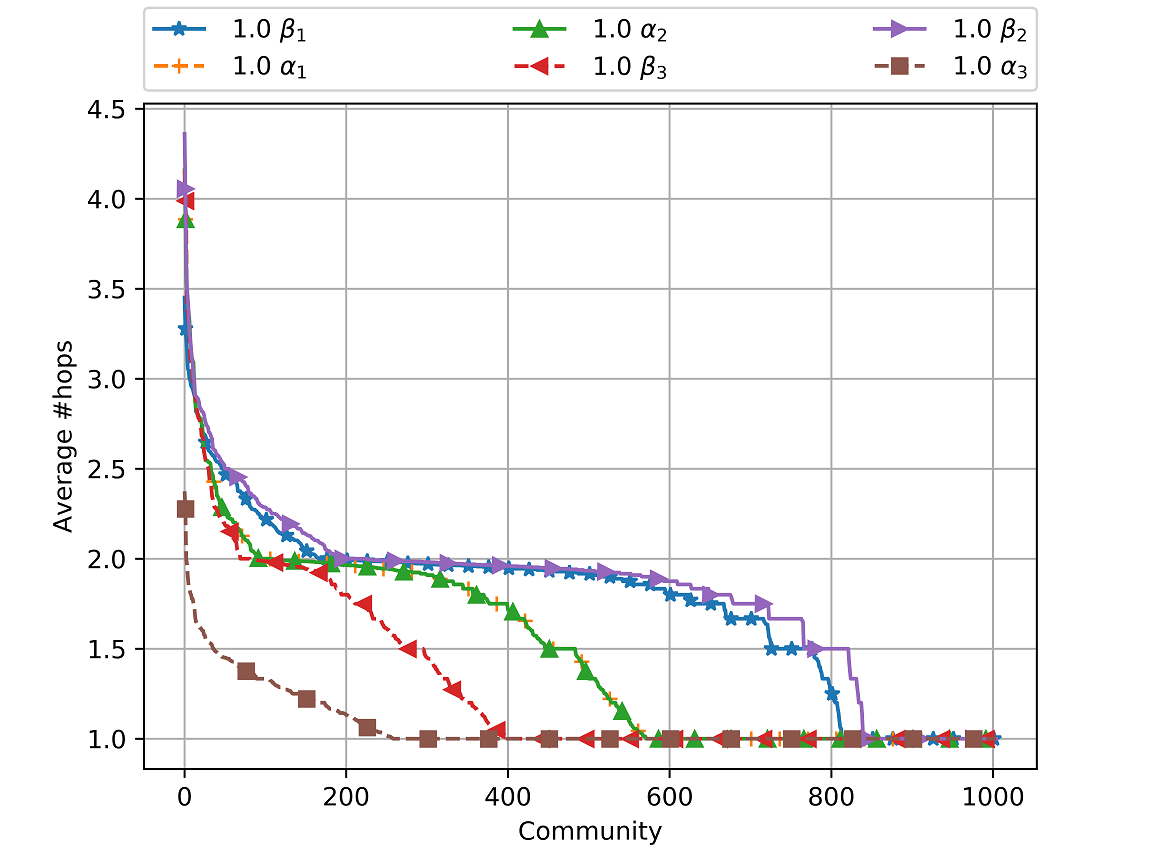}
\caption{Average number of hops-to-leader against community in decreasing order for Phase One's communities.}
\label{fig:dual-hops-phase-one}
\end{figure}

Moving on, we further evaluate this derivation by producing a visualization of the average number of hops-to-leader for each community versus community size.
Figure~\ref{fig:avg-hops} presents this with respect to natural geographical zones of guifi.net in the left, with our algorithm's results on the right side.
Our algorithm led to an overall reduction in the number of hops, in particular for smaller and more frequent communities.
Figure~\ref{fig:avg-hops-log} highlights interesting tendencies with regard to the impact of absolute heuristic weights and their influence on the average number of hops.
In particular, we achieve this by isolating the range of community sizes to a maximum size of 250 members.
Plotting these ranges over a logarithmic scale, it can be seen that the contained communities exhibit a lower number of hops.
This tendency is particularly manifested with heuristics $\alpha_{1}$ and $\beta_{3}$ (betweenness centrality and computational class of the node, respectively).
We extrapolate from this finding that the fixed-region geographical definition of guifi.net may be too rigid and that it may in fact provide a user experience which is probably below-optimal regarding typical services offered in CNMCs.
Usage of the Phase One technique shows promise with respect to optimizing the length of the path taken from each community's node to the community leader, a sure benefit for many services.

%%%%%%%%%%%%%%%%%%%%%%%%%%%%%%%%%%%%%%%%%%%%%%%%
%%%%%%%%%%%%%%%%%%%%%%%%%%%%%%%%%%%%%%%%%%%%%%%% EVALUATION - PHASE TWO
%%%%%%%%%%%%%%%%%%%%%%%%%%%%%%%%%%%%%%%%%%%%%%%%
\textbf{Phase Two: Leader Election Results.}
It is relevant to note that after Phase Two of our algorithm, the application of heuristics over the propagation-based node sets (right side) yielded more outliers than the geographical zones (left side).
While there were more outliers in the results of Phase One of our algorithm, lower values were achieved when compared to the geographical node groups.
We presented obtained results evaluated under different criteria.
Our focus is not on producing a one-size-fits-all hierarchy of heuristics: other real-world scenarios upon which to test our algorithm will have specific objective functions, bound by application needs.
The results are promising as they highlight that our algorithm is a valid alternative to traditional computational approaches to optimizing responsibility assignment to network nodes.
We present Figures~\ref{fig:dual-hops-geo} and~\ref{fig:dual-hops-phase-one}, which depict the number of average hops-to-leader in decreasing order.
Orthogonally to node group definitions, the tendencies in the influence of the heuristics remain valid, with the same patterns appearing for each of the cases.
It is interesting to note that, for the right side (based on Phase One of our algorithm), heuristics $\alpha_{2}$ and $\beta_{3}$ produced greater differences between them.
Accounting for the computational class of nodes in the case of the right side led to a lower number of hops-to-leader compared to simply electing leaders based on centrality.

\begin{figure}
\centering
\includegraphics[width=0.95\linewidth]{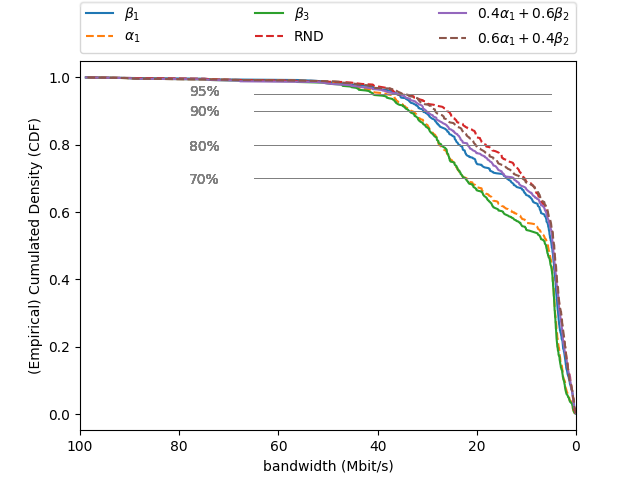}
\caption{Zone-based bandwidth ECDF.}
\label{fig:zones-bw-ecdf}
\end{figure}

\begin{figure}
\centering
\includegraphics[width=0.95\linewidth]{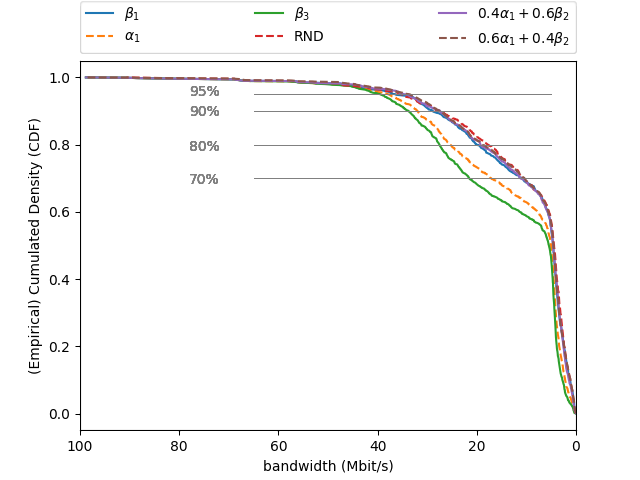}
\caption{Community-based bandwidth ECDF.}
\label{fig:unscored-bw-ecdf}
\end{figure}

\begin{figure}
\centering
\includegraphics[width=0.95\linewidth]{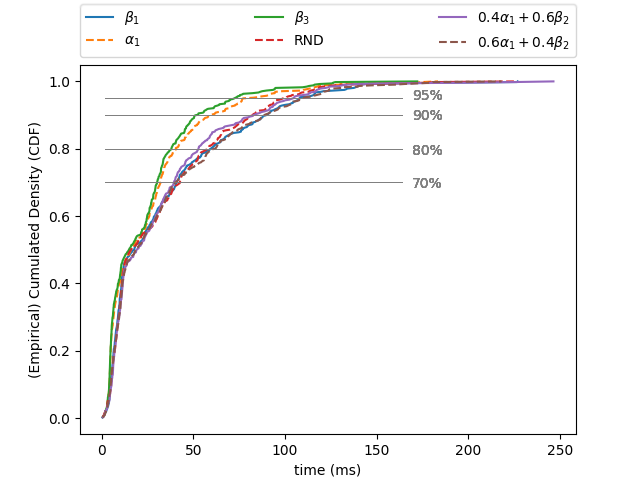}
\caption{Zone-based round-trip time ECDF.}
\label{fig:zones-rtt-ecdf}
\end{figure}

\begin{figure}
\centering
\includegraphics[width=0.95\linewidth]{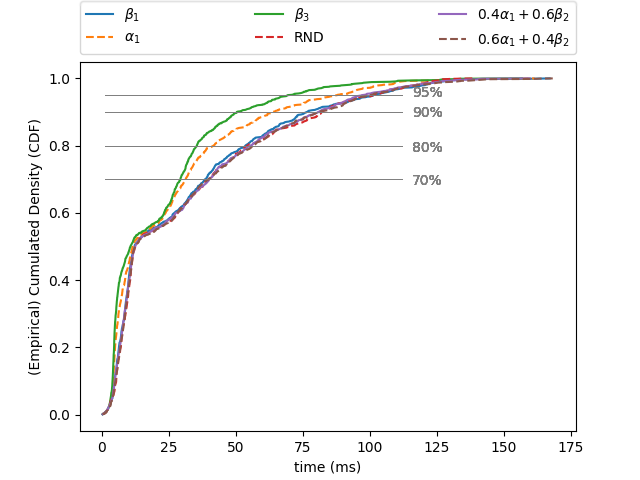}
\caption{Community-based round-trip time ECDF.}
\label{fig:unscored-rtt-ecdf}
\end{figure}

%%%%%%%%%%%%%%%%%%%%%%%%%%%%%%%%%%%%%%%%%%%%%%%%
%%%%%%%%%%%%%%%%%%%%%%%%%%%%%%%%%%%%%%%%%%%%%%%% EVALUATION - SLA ASSESSMENT
%%%%%%%%%%%%%%%%%%%%%%%%%%%%%%%%%%%%%%%%%%%%%%%%
\textbf{SLA Assessment.}
We also evaluate the quality of leaders in the context of the sampling performed for the QMP network.
Namely, we modeled round-trip time (RTT) in milliseconds and bandwidth B/W in Mbit/s distributions based on around 70,000 samples (of bandwidth and round-trip time) obtained from QMP in guifi.net over a period of seven days.
These two features are relevant to types of SLAs inherent to services such as (RTT) web caching, web content requests, NoSQL cloud storage as well as (B/W) streaming and file download services.
From these two features, we modeled their distribution and simulated their values for all of the guifi.net network snapshot mentioned earlier.
Figures~\ref{fig:zones-bw-ecdf}, ~\ref{fig:unscored-bw-ecdf}, ~\ref{fig:zones-rtt-ecdf}, and~\ref{fig:unscored-rtt-ecdf} were produced using the \texttt{Python statsmodel} package~\cite{seabold2010statsmodels}, which has a set  of utilities to automate statistical processing tasks.

Figures~\ref{fig:zones-bw-ecdf} and~\ref{fig:unscored-bw-ecdf} show the empirical cumulative distribution function of the bandwidth for the original guifi.net zones and for communities produced by Phase One of our algorithm, respectively.
Interestingly, the bandwidth interval for [10; 30] Mbit/s in Figure~\ref{fig:zones-bw-ecdf} shows that Phase Two of our algorithm (the stage of leader election within a community -- in this test case there is a one-to-one mapping between guifi.net zones and communities) fared better by using singular heuristics for electing the leader.
That is, heuristics $\beta_{3}$ (computational class) and $\alpha_{1}$ (betweenness centrality) produced, on average, more efficient bandwidth paths from a community's nodes to their leader.
This tendency was also reproduced in the execution of Phase Two of our algorithm after Phase One (custom communities generated by Algorithm~\ref{alg:one} instead of one-to-one mapping to guifi.net's original zones), which can be seen in Figure~\ref{fig:unscored-bw-ecdf}.

For the round-trip time, the same two heuristic weights fared better than the others as well.
All combinations seemingly max out (in terms of cumulative distribution) at around 125 milliseconds.
However, up to about 100 milliseconds, heuristics $\beta_{3}$ (computational class) and $\alpha_{1}$ (betweenness centrality) produced lower round-trip time.
This occurs for the zone-based communities in Figure~\ref{fig:zones-rtt-ecdf} and also the communities resulting from Phase One of our algorithm, as illustrated in Figure~\ref{fig:unscored-rtt-ecdf}.
Curiously, we observe, as far as round-trip time is concerned, that the random leader election yielded practically the same results as the combined usage of betweenness centrality $\alpha_{1}$ and latency $\beta_{2}$ (alternating between $0.4\alpha_{1}+0.6\beta_{2}$ and $0.6\alpha_{1}+0.4\beta_{2}$).
We did not perform an exhaustive analysis of all possible combinations of heuristics and their weights.
The combinations we present herein are relevant in terms of what the heuristics represent.
Bandwidth samples were modeled as a four-parameter Kappa distribution, while round-trip time was modeled as a generalized extreme value (GEV) distribution.
It is relevant to say that the empirical distributions of these two metrics exhibited a considerable degree of independence.
In fact, $corr$($BW, R$)$= -0.134$ for the sampled values, which means they appear to be only slightly inversely related.
We assumed them to be independent with respect to results.

\textbf{Summary.}
The method we present is inherently parallel and distributed, a break from traditionally-centralized often exhaustive optimization-driven solutions, opening possibilities for scalability.
Phase Two of our algorithm was designed to be distributed with the purpose of executing concurrently among all communities.
This implies that the computational time of this phase has an upper bound associated to the slowest-computing community.
As far as the authors are aware, this work is the first that attempts to optimize service placement by defining communities using an analysis based purely on network theory and distributed graph processing.
The guifi.net telecommunications network is one upon which different research projects have been executed~\cite{6379103,6379139}.

%%%%%%%%%%%%%%%%%%%%%%%%%%%%%%%%%%%%%%%%%%%%%%%%
%%%%%%%%%%%%%%%%%%%%%%%%%%%%%%%%%%%%%%%%%%%%%%%% RELATED WORK
%%%%%%%%%%%%%%%%%%%%%%%%%%%%%%%%%%%%%%%%%%%%%%%%

\section{Related Work}\label{sec:related}
Herein we go over alternative approaches to Phase One and Phase Two of our solution as a whole.
We note that our work is novel, as far as we know, in the sense that it combines these two multidisciplinary phases, whose literature we analyze.

%%%%%%%%%%%%%%%%%%%%%%%%%%%%%%%%%%%%%%%%%%%%%%%%
%%%%%%%%%%%%%%%%%%%%%%%%%%%%%%%%%%%%%%%%%%%%%%%% RELATED WORK - ON COMMUNITY DETECTION
%%%%%%%%%%%%%%%%%%%%%%%%%%%%%%%%%%%%%%%%%%%%%%%%
\textbf{Community Networks.}
Different studies on guifi.net have drawn several insights: the network is not homogeneous -- rural areas have topology properties different from those of metropolitan areas, such as density; the topology observed in rural areas is not scale-free (degree distribution does not fit a power law) due to the high number of terminals connected to some nodes; removing terminal nodes (with degree one) from the graphs in rural areas, however, reveals a scale-free \textit{core-network} as in~\cite{6379139}.
On the one hand, it is necessary to be aware of the challenges inherent to service allocation in different types of networks in the context of distributed systems.
On the other hand, we highlight the existence of community detection techniques (in network theory) as a novel approach to these challenges.
In recent years, metrics have been proposed for evaluating the quality of calculated communities have emerged: the most notorious one being that of modularity.
However, focusing exclusively on modularity incurs community resolution penalties with smaller communities often not being detected.
Considering this and focusing on scalability, other methods in the literature which do not use domain-specific heuristics were devised, such as the class of label propagation algorithms~\cite{PhysRevE.79.066107,PhysRevE.76.036106}.
These algorithms are inherently parallel and work well in practice for real world networks~\cite{Boldi:2011:LLP:1963405.1963488}.

%%%%%%%%%%%%%%%%%%%%%%%%%%%%%%%%%%%%%%%%%%%%%%%%
%%%%%%%%%%%%%%%%%%%%%%%%%%%%%%%%%%%%%%%%%%%%%%%% RELATED WORK - ON SERVICE PLACEMENT
%%%%%%%%%%%%%%%%%%%%%%%%%%%%%%%%%%%%%%%%%%%%%%%%
\textbf{Service Placement.}
Typically, by monitoring all the physical and virtual resources on a system, service placement aims to balance load through the allocation, migration and replication of tasks. 
This can take place in cloud data-centers and in wireless networks that power a significant part of CNs. 
Most of the work in the data center environment, including distributed data centers, is not applicable to our case because we have a strong heterogeneity given by the limited capacity of nodes and links, as well as asymmetric quality of wireless links. 
The authors in~\cite{Veg14} introduce a service allocation algorithm that provides near-optimal overlay allocations without the need to verify the whole solution space.
They use static data from the network to identify node traits and minimize the coordination and overlay cost along a network.
The work in~\cite{Nov15} analyzes network topology and service dependencies, and 
combined with set of system constraints determines the placement of services within the wireless network.
The authors use a multi-layer model to represent a service-based system embedded in a network topology and then apply an optimization algorithm to this model to find where best to place or reposition the services as the network topology and workload on the services changes. 
%Mennan added this paragraph

In distributed micro-cloud environment (i.e., similar to our case), the work of Elmroth \cite{Erik2016} takes into account rapid user mobility and resource cost when placing applications in Mobile Cloud Networks (MCN). A recent work of Tantawi \cite{Tantawi2016} uses biased statistical sampling methods for cloud workload placement. Regarding the service placement through migration, the authors in \cite{Wang2017} study the dynamic service migration problem in mobile edge-clouds that host cloud-based services at the network edge. They formulate a sequential decision making problem for service migration using the framework of Markov Decision Process (MDP) and illustrate the effectiveness of their approach by simulation using real-world mobility traces of taxis in San Francisco. 
As a whole, mostly, service placement approaches are predominantly based on resource  (CPU, memory) and node availability, and when they are network-aware, they are able just to employ static network information or at most process historical network data for availability predictions.
Moreover, they are batch-oriented and execute sequentially in centralized settings and therefore cannot scale to larger network sizes, number of services, or greater network dynamism.
Our approach is the first, to the best of our knowledge, that is dynamic, parallel and distributed, and therefore able scale seamlessly, by employing distributed graph processing systems, such as the \texttt{Gelly} library of \texttt{Apache Flink}.
Thus, we are able to continually monitor service quality and perform service placement decisions continually/incrementally based on data gathered from the network (e.g., graph-servers in guifi.net).
%%%%%%%%%%%%%%%%%%%%%%%%%%%%%%%%%%%%%%%%%%%%%%%%
%%%%%%%%%%%%%%%%%%%%%%%%%%%%%%%%%%%%%%%%%%%%%%%% CONCLUSION
%%%%%%%%%%%%%%%%%%%%%%%%%%%%%%%%%%%%%%%%%%%%%%%%
\section{Conclusion}\label{sec:conclusion}
In this paper, we presented a novel take on the processing steps that underlie service placement, a multi-objective problem.
Compared to traditional system techniques (which, as far as we know, have not seen developments regarding parallel implementations and scalability with network size), our algorithm is expressed purely over state-of-the-art graph techniques which have inherent parallelism.
This makes our algorithm a very competitive alternative, able to scale for networks which are orders of magnitude greater, when compared to other traditional techniques in the field.

\begin{acks}
\footnotesize{This work was partly supported by the Portuguese government through FCT -- Funda{\c c}{\~a}o para a Ci{\^e}ncia e Tecnologia, under projects PTDC/EEI-SCR/6945/2014 and UID/CEC/500021/2013, by the ERDF through COMPETE 2020 Programme, within project POCI-01-0145-FEDER-016883, by the European H2020 project LightKone (H2020-732505), and by the Spanish government under contract TIN2016-77836-C2-2-R.}\normalsize
\end{acks}

\bibliographystyle{ACM-Reference-Format}
\bibliography{references} 

\end{document}